%% file: flames.tex
\documentclass[twocolumn]{aastex62}

\usepackage{amsmath}
\usepackage{units}

\input{macros.tex}

\input{nuclides.tex}

\newlength{\apjcolwidth}
\begin{document}

\author[0000-0002-4870-8855]{Josiah Schwab}
\affiliation{Department of Astronomy and Astrophysics, University of California, Santa Cruz, CA 95064, USA}
\correspondingauthor{Josiah Schwab}
\email{jwschwab@ucsc.edu}

\author[0000-0003-3441-7624]{R. Farmer}
\affiliation{Anton Pannenkoek Institute for Astronomy and GRAPPA, University of Amsterdam, NL-1090 GE Amsterdam, The Netherlands}
\affiliation{Center for Astrophysics, Harvard-Smithsonian, 60 Garden Street, Cambridge, MA 02138, USA.}

\author[0000-0002-0474-159X]{F.X.~Timmes}
\affiliation{School of Earth and Space Exploration, Arizona State University, Tempe, AZ 85287, USA}
\affiliation{Joint Institute for Nuclear Astrophysics - Center for the Evolution of the Elements, USA}

\title{Laminar Flame Speeds in Degenerate Oxygen-Neon Mixtures}

\begin{abstract}
  The collapse of degenerate oxygen-neon cores (i.e., electron-capture
  supernovae or accretion-induced collapse) proceeds through a phase
  in which a deflagration wave (``flame'') forms at or near the
  center and propagates through the star.  In models, the assumed speed of
  this flame influences whether this process leads to an explosion or
  to the formation of a neutron star.  We calculate the laminar flame
  speeds in degenerate oxygen-neon mixtures with compositions motivated by
  detailed stellar evolution models.  These mixtures include trace
  amounts of carbon and have a lower electron fraction than those
  considered in previous work.  We find that trace carbon has little
  effect on the flame speeds, but that material with electron fraction
  $\Ye \approx 0.48-0.49$ has laminar flame
  speeds that are $\approx 2$ times faster than those at $\Ye = 0.5$.  We
  provide tabulated flame speeds and a corresponding fitting function
  so that the impact of this difference can be assessed via
  full star hydrodynamical simulations of the collapse process.
\end{abstract}

\keywords{White dwarf stars (1799); Degenerate matter (367); Oxygen burning (1193)}

\section{Introduction}

Degenerate oxygen-neon (ONe) cores with masses near the Chandrasekhar mass
can form in the evolution of $\approx \unit[8-10]{\Msun}$ single stars
\citep[e.g.,][]{Miyaji1980, Miyaji1987}, in interacting binary systems
with varying degrees of envelope stripping
\citep[e.g.,][]{Tauris2015b, Poelarends2017}, in binary systems with
an accreting ONe white dwarf (WD) \citep[e.g.,][]{Canal1976,
  Nomoto1991}, or as the result of the merger of two WDs
\citep[e.g.,][]{Saio1985b, Brooks2017d}.
When the core reaches a central density $\approx \unit[10^{10}]{\gcc}$,
exothermic electron captures on \neon[20] occur and lead to the
initiation of a deflagration wave (``flame'') that propagates outward.
The competition between the energy release from this flame and the
electron-capture reactions on its ashes determines whether this leads
to an explosion (resulting in partial or total disruption of the star)
or implosion (resulting in the formation of a neutron star (NS)).

This situation has long been known to be finely balanced
\citep[e.g.,][]{Nomoto1991, Canal1992}, though the general conclusion
by the end of the 1990s was in favor of collapse to a NS.  Recent
multidimensional simulations have reiterated that the outcome is
sensitive to modeling choices and reopened the possibility that at
least some cases may lead to a thermonuclear explosion (possibly also leaving a low-mass bound remnant) instead of
collapse to an NS \citep{Jones2016c, Leung2019b, Jones2019a}.  One of
the key ingredients in this modeling is the speed at which the flame
propagates.

\citet{Timmes1992}, hereafter \tw, calculated the physical properties
of conductively-propagated laminar burning fronts in high-density,
degenerate carbon-oxygen (CO) and ONe mixtures.  We repeat a similar
set of calculations using Modules for Experiments in Stellar
Astrophysics \citep[\MESA;][]{Paxton2011, Paxton2013, Paxton2015,
  Paxton2018, Paxton2019}, but extend these results to a wider range
of compositions motivated by expectations from detailed models of the
internal composition of ONe WDs \citep[e.g.,][]{GarciaBerro1997b, Iben1997, Siess2006}.
Section~\ref{sec:methods} describes the methods we use to calculate
the laminar flame speeds.  Section~\ref{sec:tw92} reproduces the \tw\
results in both CO and ONe mixtures.  We then focus on the laminar
flame speeds in ONe mixtures under different conditions.  
In Section~\ref{sec:trace-c} we show
how the flame speeds  are only mildly affected by the presence of small amounts of
\carbon[12], but in Section~\ref{sec:low-ye} demonstrate the
significant influence of the electron fraction (\Ye) of the material.
Section~\ref{sec:tables} provides
tabulated flame speeds and a corresponding fitting function.
Section~\ref{sec:conclusions} briefly describes the implications for models of
the collapse of ONe cores.

\section{Methods}
\label{sec:methods}

We use \MESA\ revision r12115 \citep{MESA}.  The input files necessary
to reproduce our work are publicly available\footnote{\url{https://doi.org/10.5281/zenodo.3537874}}
and an illustration of this capability has been included in the test
suite case \texttt{conductive\_flame}.

We create an initial, spatially-uniform \MESA\ model with a temperature
$\Tb = \unit[3\times10^8]{K}$, specified density
$\rho_9 = \rho / (10^9\,\gcc)$, and a specified unburned composition.
These properties characterize the cold material into which the flame
will propagate.  So long as the upstream temperature is much less than
the downstream (post-burn) temperature of $\approx \unit[10^{10}]{K}$, the
temperature jump across the flame is approximately the same and 
the initial temperature does not play an important role.
The total mass $M$ defines the size of the (Lagrangian) computational
domain.  Because the flame width $\lambda$ varies substantially with
density, our domain size must as well.  We always ensure that
$M/(\rho \lambda^3) \gg 1$, but typically choose this ratio to be
$\sim 100$ to limit the computational cost.  This also implies that
$\lambda \ll r$, so the flame is effectively planar.  In practice,
$M \sim \unit[10 - 10^5]{g}$.  The small size of the domain implies
that the pressure gradient due to gravity is negligible.  The inner
boundary is at $r = 0$.  The outer boundary has a fixed temperature
$\Tb$ and a fixed pressure equal to the initial pressure of the material.

We then insert a hot spot at the center with fractional size in mass
$\qs = 3\times10^{-4}$ at a temperature $\Ts = \unit[8\times10^9]{K}$
(for CO mixtures) or  $\Ts = \unit[10\times10^9]{K}$ (for ONe mixtures).
This hot spot should have a size of order the flame width and a
temperature of order the post-burn temperature to ensure a steady-state flame.
If the hot spot is too small the flame will die.
If the hot spot is too large the flame will exhibit a de-accelerating transient.
With a well chosen spot size much smaller than the domain, once
the flame has propagated a few flame lengths, the initial condition
will be effectively erased.

\begin{figure} [!htb]
  \includegraphics[width=\columnwidth]{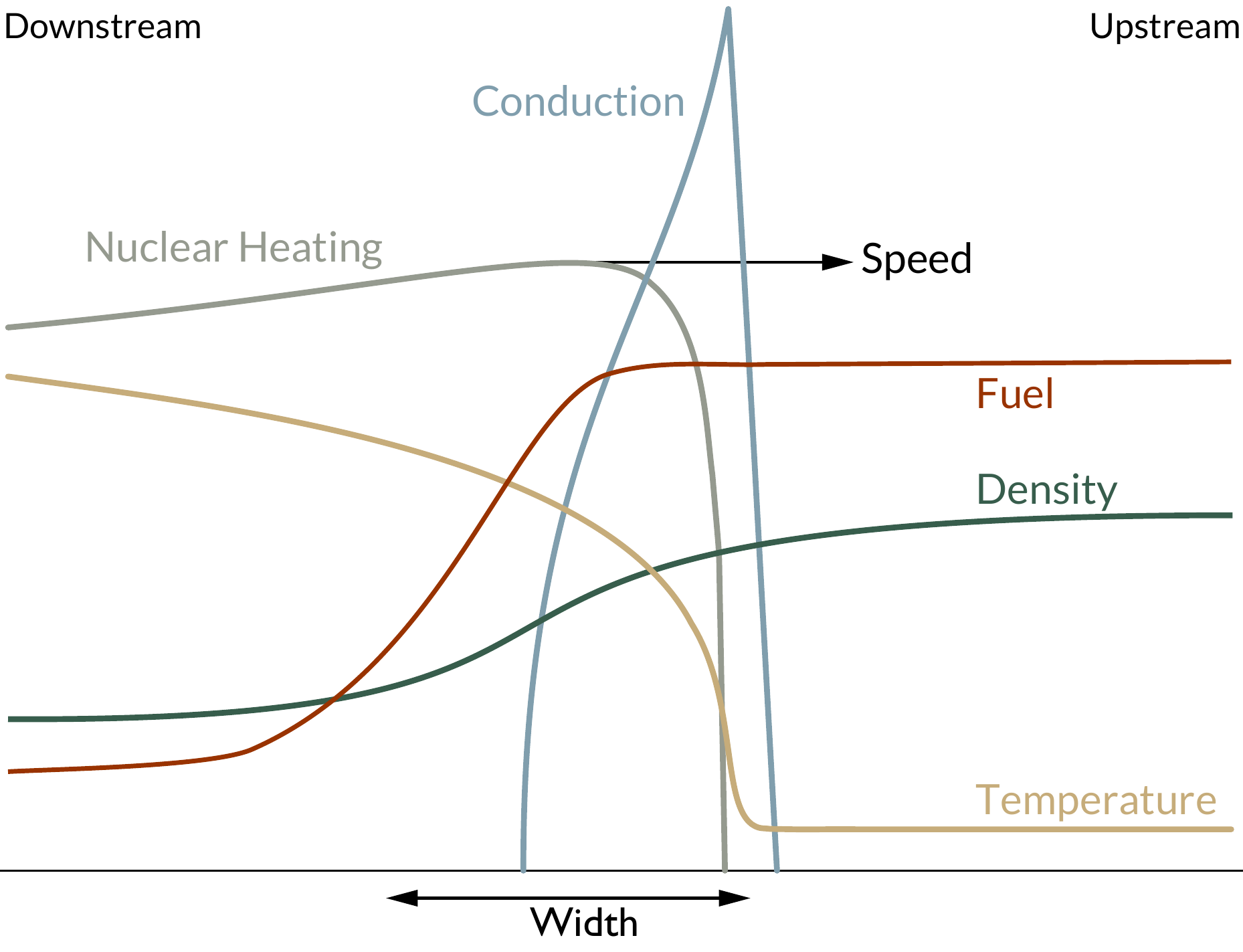}
  \caption{Schematic of a steady-state flame. The fuel in the upstream material 
           is initially heated by conduction until the temperature becomes large
           enough to ignite nuclear reactions. At the critical temperature, 
           the energy generation rate equals the energy conduction term (heating equals cooling).
           The downstream material burns to its nuclear statistical equilibrium state.
           The entire structure, approximately isobaric, propagates into the upstream fuel with a unique speed and width.}
  \label{fig:flame}
\end{figure}

We define the location of the flame to be the location of the maximal
rate of nuclear energy release in the domain (i.e., peak nuclear heating, see Figure~\ref{fig:flame}).  We evolve the model
until the flame has propagated through 90\% of the domain and then
extract the steady-state flame properties.  By repeating this process for different
initial conditions, we calculate the laminar flame speed as a function
of $\rho_9$ and composition. 
In Appendix~\ref{sec:convergence}, we
demonstrate that our results are insensitive
to the details of the initial conditions
and are numerically converged.

\subsection{Microphysics}
\label{sec:micro}

As discussed in \tw, the flame will have a width such that the
diffusion timescale across it is comparable to the timescale at which
nuclear reactions heat the material.  This argument leads to an
estimate of the flame speed,
\begin{equation}
  v_{\rm flame} \approx \left(\frac{D_{\rm th}\epsnuc}{e}\right)^{1/2}~,
  \label{eq:vflame}
\end{equation}
where $D_{\rm th}$ is the thermal diffusion coefficient, $\epsnuc$ is
a characteristic specific rate of energy generation from nuclear
reactions, and $e$ is a characteristic specific energy.  Therefore,
the speed of the flame is set by the energy generation rate as
determined from the nuclear network and the assumed thermal transport
properties of the degenerate plasma.

\subsubsection{Transport properties}

\begin{figure*}
  \includegraphics[height=6.6cm]{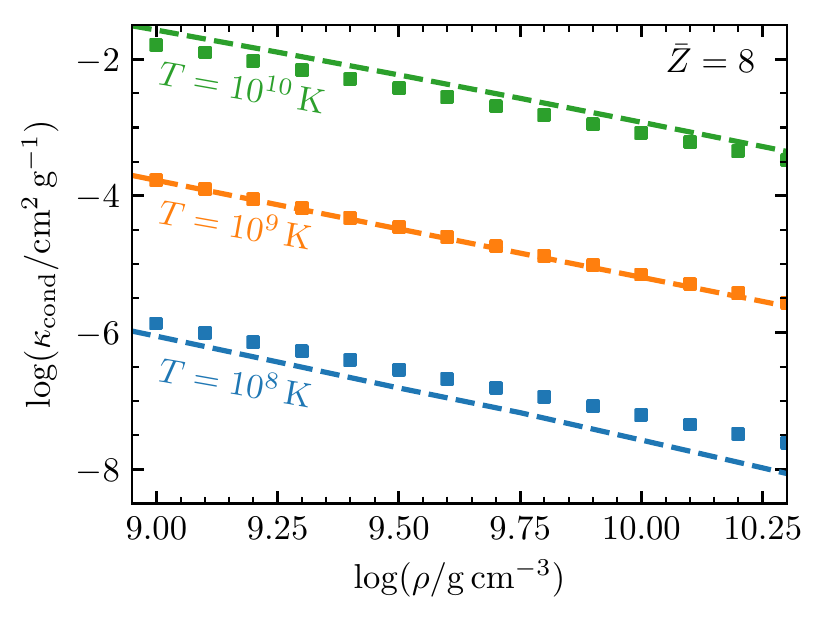}
  \hfill
  \includegraphics[height=6.6cm]{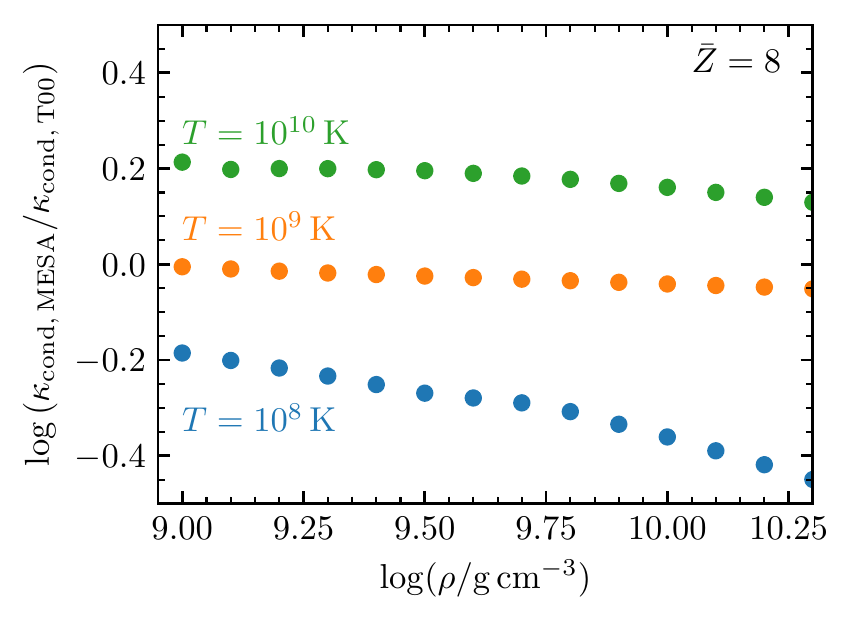}
  \caption{Comparison of conductive opacity for pure \oxygen[16]. The square points show the values using the routines from T00, which are similar those used in \tw.  The dashed line shows the values adopted in \MESA. The right panel shows the relative difference between the two sets of values.}
  \label{fig:kap-cond}
\end{figure*}

In \MESA\, the thermal conductivity of the degenerate electrons,
accounted for via a conductive opacity
$(\kappacond \propto D_{\rm th}^{-1})$, comes from tables privately
communicated by A.Y.~Potekhin \citep[see Section A.3
in][]{Paxton2013}.  \tw\ describe in detail their implementation of
the transport properties; a substantially similar approach is adopted
in \citet{Timmes2000a}, hereafter T00.  The source code for the
transport properties assumed in these works is publicly
available.%
\footnote{\url{http://cococubed.asu.edu/code_pages/kap.shtml}}

Figure~\ref{fig:kap-cond} compares the conductive opacities over the
range of temperatures and densities considered in this paper.  The
left panel shows the T00 and MESA values of $\kappacond$ and the right
panel their relative difference.  Both sources show similar density
scalings, but the T00 values scale less steeply with temperature such
that while the values agree at $\unit[10^9]{K}$, $\kappacond$ is
higher by a factor $\approx 2$ at $\unit[10^8]{K}$ and lower by a
factor $\approx 1.5$ at $\unit[10^{10}]{K}$.  Since these variations
are not systematically in the same direction, their effect is
difficult to estimate, but given the scaling in
Equation~(\ref{eq:vflame}), variations in $\kappa_{\rm cond}$ at this
level correspond to $\approx 30\%$ variations in the flame speed.

\subsubsection{Nuclear reaction rates}

\begin{figure}
  \fig{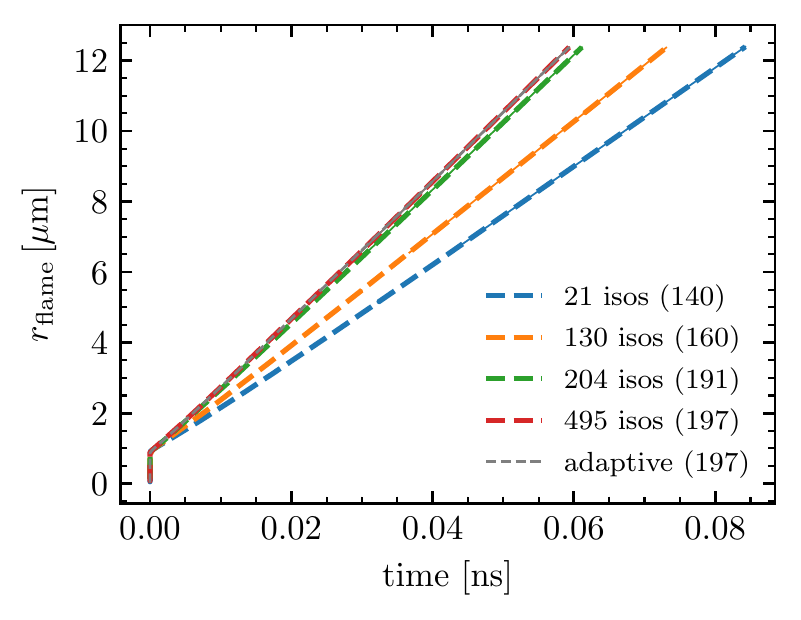}{\columnwidth}{(a) CO mixture with $\Xc = 0.5$ at $\rho_9 = 6$.}
  \fig{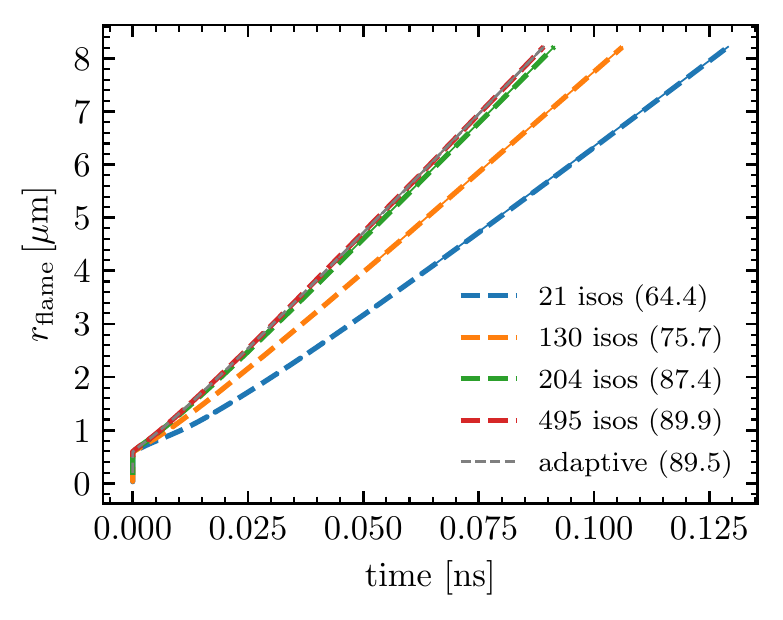}{\columnwidth}{(b) ONe mixture with $\Xo = 0.6$ at $\rho_9 = 10$.}
  \caption{Effect of nuclear network size.  For each calculation, the
    flame location is shown as a function of time.  The legend lists
    the number of isotopes (``isos'') used and the flame speed (in
    \kms) is indicated in parenthesis.  The thin solid line underlying
    the dashed line indicates the portion of the curve used to extract
    the flame speed.
    (The ``adaptive'' and ``495 isos'' curves overlap to within the line width.)}
  \label{fig:net-size}
\end{figure}

The currently applicable default inputs for nuclear reaction rates are
described in Appendix A.2 of \citet{Paxton2019}.  Rates are taken from
a combination of NACRE \citep{angulo99} and the Joint Institute for
Nuclear Astrophysics REACLIB library (default version, dated
2017-10-20) \citep{Cyburt2010}.  The \MESA\ screening corrections are
from \citet{Chugunov2007}, which includes a physical parameterization
for the intermediate screening regime and reduces to the familiar weak
\citep{Dewitt1973, Graboske1973} and strong \citep{Alastuey1978,
  Itoh1979} limits at small and large values of the plasma coupling
parameter.

Relatively large nuclear networks are required to fully capture the
energy generation rate in neutron-rich compositions at these
thermodynamic conditions. \tw\ illustrate the increase in flame
speed with increasing network size (their Table 5) and adopt a 130
isotope network.  We perform a similar exercise, using 3 built-in
\MESA\ networks (\texttt{approx21}, \texttt{mesa\_204},
\texttt{mesa\_495}), a network constructed with the same elements as
the \tw\ 130 isotope network (see their Table 1), and also an adaptive
network that automatically adds and removes isotopes and which settles
in at around 320 isotopes.  Figure~\ref{fig:net-size} shows the flame
location as a function of time for a set of runs for a fiducial CO
mixture (panel a) and a fiducial ONe mixture (panel b).  Networks of
more than 200 isotopes appear to be required before network size no
longer makes an appreciable difference in the flame speed.  This
result is consistent with \citet{Chamulak2007}, hereafter C07, who
found that for flames in CO mixtures a 430 isotope network gave speeds
up to $\approx 25\%$ greater than a 130 isotope network.  We run with
495 isotopes unless otherwise stated.

The JINA REACLIB polynomial fits to the reaction rate data end at
$\unit[10^{10}]{K}$ as do the tabulated partition functions used to
calculate the reverse rates and ensure detailed balance.  Above
$\unit[10^{10}]{K}$, \MESA\ fixes the rates to be their
$T = \unit[10^{10}]{K}$ values.  In some cases, especially for the ONe
flames, the temperature exceeds $\unit[10^{10}]{K}$ and the peak in
$\epsnuc$ occurs near the temperature threshold.  If the \MESA\
treatment underestimates the true peak of $\epsnuc$, then this can
lead to an underestimate of the flame speed.  (For example, if we
truncate the rates at $T = \unit[8\times10^{9}]{K}$, the ONe flame
in Figure~\ref{fig:net-size} has a speed of \unit[82]{\kms}, a
$\approx 10\%$ reduction.)  However, we have physical reasons to
expect that this effect is not particularly large.  By
$\approx \unit[1.2\times10^{10}]{K}$ photodisintegration is strong
enough to decompose nuclei into neutrons, protons, and alpha
particles.  This is an endothermic process, meaning there is an upper
limit to how much more positive $\epsnuc$ can be achieved beyond the
place where \MESA\ truncates the rates.

Our results depend slightly on our adopted rate sources.  If we use
pure JINA REACLIB defaults (eliminating NACRE), the flame speeds
increase.  For the calculation shown in Figure~\ref{fig:net-size}, the
result with the 495 isotope net and only the JINA rates is
\replaced{
\unit[94.1]{\kms} for the CO case and \unit[207]{\kms} for the ONe
case.}{
\unit[207]{\kms} for the CO case and \unit[94.1]{\kms} for the ONe
case.}
These represent an approximately 5\% speed up.

Thus there is some systematic uncertainty from nuclear reaction rates in our results which is difficult
to characterize, but seems unlikely to be smaller than $\approx 10\%$.
We note that both the above caveats result in even faster flame speeds
than the ones we will report.
The open and reproducible nature of our work allows this problem to be
  easily revisited, enabling the impact of future experimental and theoretical progress in the relevant reaction rates to be quickly assessed.

\section{Comparison with past work}
\label{sec:tw92}

\begin{figure}
  \includegraphics[width=\columnwidth]{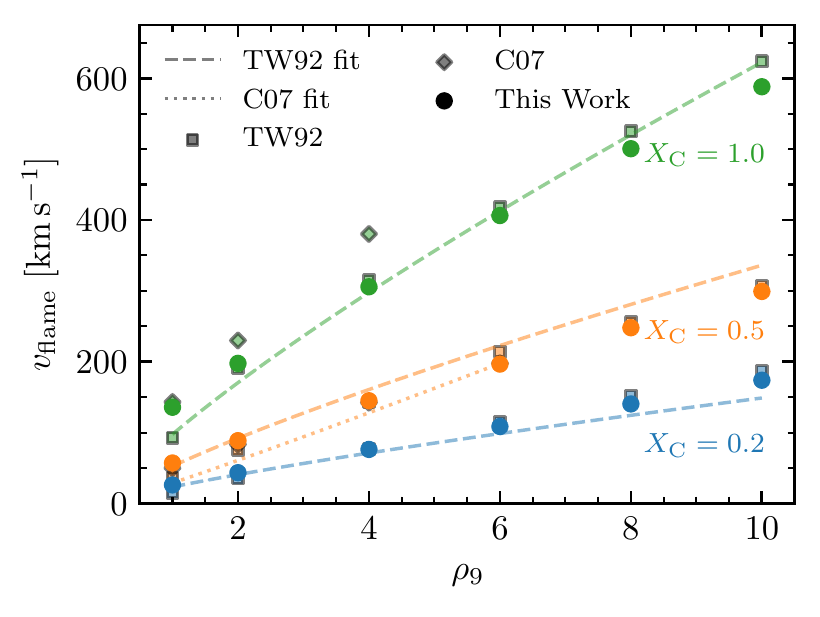}
  \caption{Comparison with \tw\ for CO mixtures.  We compare our
    results (solid circles) to both their tabulated speeds (lighter, outlined symbols) and provided fitting function (line).  We do the same for
    the subset of conditions that have data from C07
    \citep{Chamulak2007}.  This is tabulated points and a fit for
    $\Xc = 0.5$, $\rho_9 \le 6$ and points for
    $\Xc = 1.0$, $\rho_9 \le 4$.}
  \label{fig:tw92-co}
\end{figure}

First, we consider CO mixtures.  Following \tw\, we select a
\carbon[12] mass fraction and put the remainder in \oxygen[16].
Figure~\ref{fig:tw92-co} compares our results with those of \tw.
Qualitatively, the agreement is good, and we reproduce the trends with
$\rho_9$ and $\Xc$.  Quantitatively, above $\rho_9 = 4$, our results
are $\approx 5-10\%$ slower, while below $\rho_9 = 4$, our results are
faster, up to $\approx 40\%$ at $\rho_9 = 1$.

Figure~\ref{fig:tw92-co} also compares the subset of our results that
overlap with C07. We agree well at $\Xc = 0.5$.  (It
is difficult to see the symbols as they overlap.)
We note, as do C07, that their
fitting function does not appear to do a good job of matching their
tabulated results.  Our results are slower at $\Xc = 1$,
though we note the C07 values also disagree with \tw\ and that
the primary focus of C07 was on $\Xc = 0.3 - 0.7$.

\begin{figure}
  \includegraphics[width=\columnwidth]{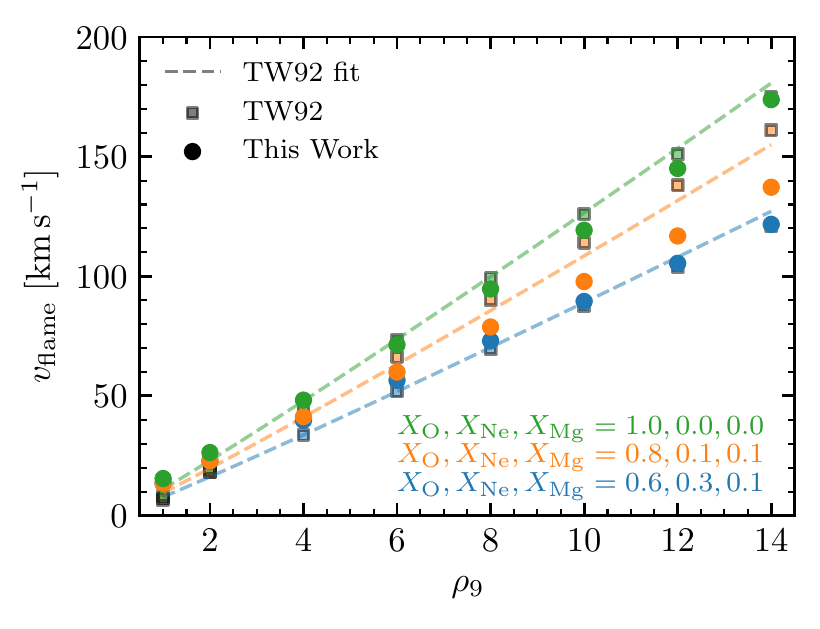}
  \caption{Comparison with \tw\ for O/Ne/Mg mixtures.  We compare our
    results (solid circles) to both their tabulated speeds (lighter,
    outlined symbols) and provided fitting function (line).
    Composition labels appear in the same vertical order as their
    corresponding lines.}
  \label{fig:tw92-one}
\end{figure}

Next, we consider ONe mixtures.  Following \tw, we first select a
mass fraction \Xo\ of \oxygen[16].  When the mixture is not pure \oxygen[16], we
also choose a mass fraction 0.1 of \magnesium[24].  The remainder is
\neon[20].  Figure~\ref{fig:tw92-one} compares our results with those
of \tw.
There is qualitative agreement, with a trend (as in the CO case) that
our flame speeds are faster than \tw\ below $\rho_9 = 4$, up to
$\approx 50\%$ at $\rho_9 = 1$.  Above $\rho_9 = 4$ and for
$\Xo = 0.6$ and $\Xo = 1.0$, the agreement is within $\approx 10\%$ of
\tw.  For $\Xo = 0.8$, the agreement is somewhat worse and the speeds
are systematically $\approx 15\%$ lower above $\rho_9 = 6$.

This section demonstrates that our speeds are generally in good
agreement ($\approx 10\%$) with the results of past work.  Relative to
\tw, our calculations adopt a larger nuclear network (leading to
faster flames), but have slightly higher conductive opacities (leading
to slower flames).  In the end, these effects may offset somewhat.  We
have no reason to expect exact agreement with \tw\ as this is not an
identical calculation.

\section{Influence of trace carbon}
\label{sec:trace-c}

\begin{figure}
  \includegraphics[width=\columnwidth]{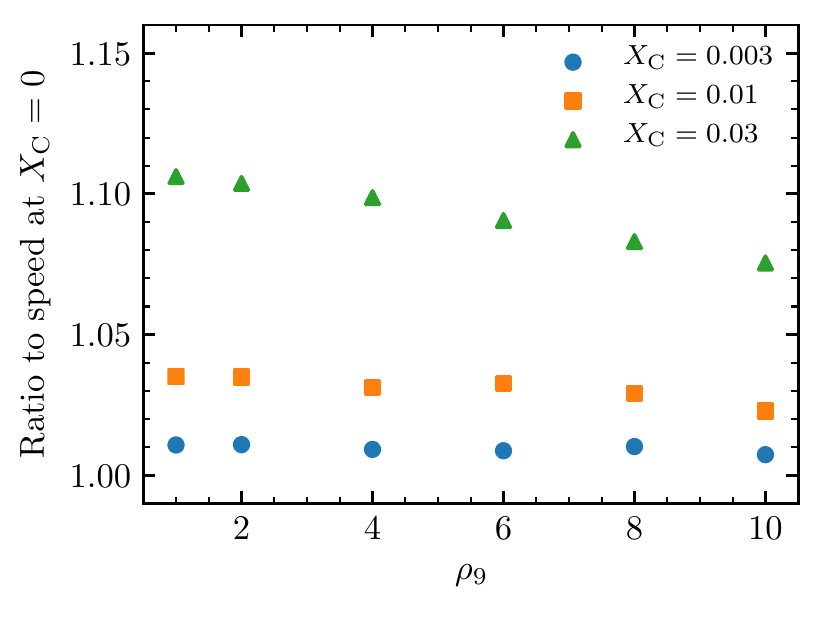}
  \caption{Effect on flame speed of trace carbon in an ONe mixture.  Each point shows
    the ratio of $v_{\rm flame}$ with the indicated mass fraction of
    \carbon[12] to an equivalent calculation with no \carbon[12].}
  \label{fig:trace_C}
\end{figure}

ONe cores are formed after off-center carbon ignition occurs
and a convectively-bounded carbon deflagration propagates to the
center \citep[e.g.,][]{Farmer2015}.  Incomplete carbon burning that
occurs as the flame approaches the center can leave residual carbon
mass fractions of up to a few percent.  \citet{Schwab2019} performed
calculations of accreting ONe WDs including the presence of this
carbon and concluded that models are unlikely to reach carbon
ignition (and subsequently oxygen ignition and the formation of the
deflagration) below the threshold density for \magnesium[24] electron
captures.  Here we explore whether, once the deflagration is ignited,
the carbon affects the flame speed.

We select the $X_{\rm O} = 0.6$, $X_{\rm Ne} = 0.3$,
$X_{\rm Mg} = 0.1$, composition used by \tw\ and add a small amount of
\carbon[12], reducing the \oxygen[16] mass fraction accordingly.
Figure~\ref{fig:trace_C} shows the ratio of this flame speed to the
carbon-free speed shown in Figure~\ref{fig:tw92-one}.  The flame speed
increases, reflecting the additional energy release from fusion of
\carbon[12] (relative to the \oxygen[16] that it replaced).  However,
for carbon mass fractions of a few percent, the flame speed increases
only by $\approx 10\%$.  Therefore, we conclude that the presence of
small amounts of carbon is unlikely to have a significant effect on
the laminar flame speeds.

\section{Influence of lower electron fraction}
\label{sec:low-ye}

\begin{figure}
  \includegraphics[width=\columnwidth]{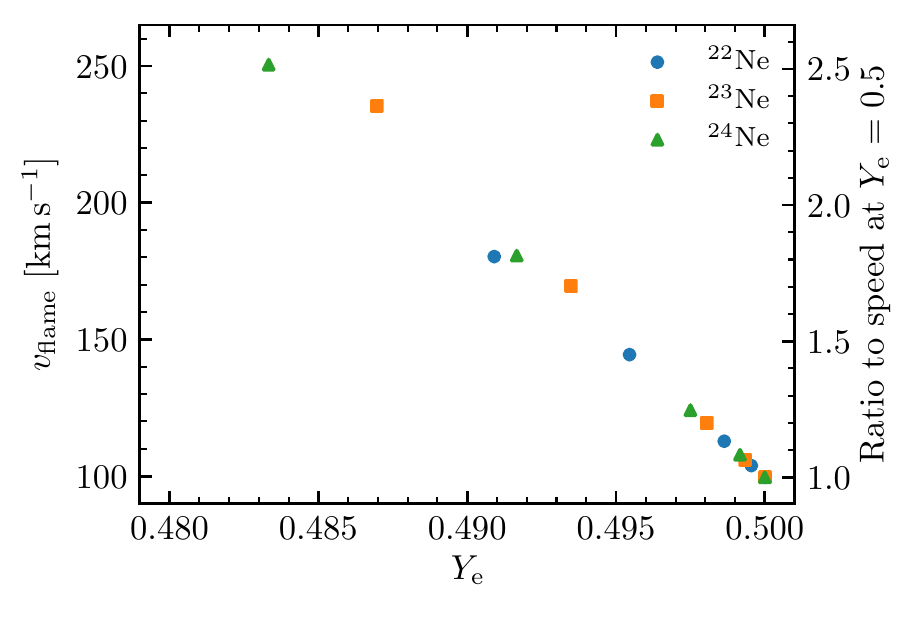}
  \caption{Effect of lower \Ye\ on flame speed in a primarily
    \oxygen[16]/\neon[20] mixture at $\rho_9 = 10$.  The different
    sequences of points achieve the $\Ye$ values by varying the mass
    fractions of the indicated Ne isotope.}
  \label{fig:lowye-ne}
\end{figure}

\begin{figure}
  \includegraphics[width=\columnwidth]{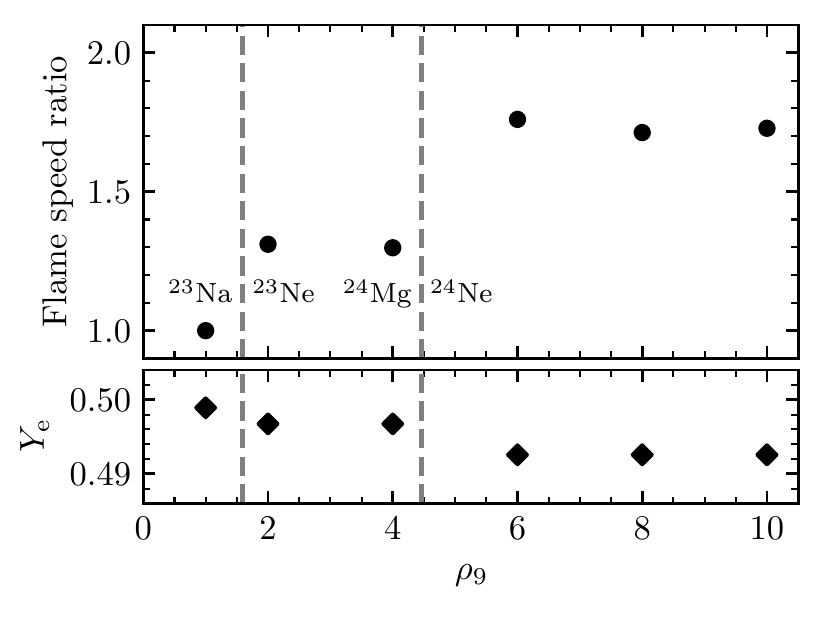}
  \caption{Effect of the density-dependent \Ye\ on flame speed.  The
    composition is $\Xo = 0.6$ and $\Xne = 0.3$ (\neon[20]) with 0.05
    each of the $A =23$ and $A=24$ isotopes.  The upper panel shows
    the ratio of the flame speed in a calculation where these were
    transformed based on density to the neutron-rich isotopes
    \neon[23] and \neon[24] to one where they remained \sodium[23] and \magnesium[24].  The dashed
    lines show the locations of these composition shifts.  The electron fraction of
    the material is indicated in the lower panel.}
  \label{fig:lowye}
\end{figure}

Detailed models of ONe WDs do not give compositions that are only
\oxygen[16], \neon[20], and \magnesium[24].  Several neutron-rich
isotopes are typically present at mass fractions of a percent or more,
meaning that material is expected to have $\Ye$ significantly below
the $\Ye = 0.5$ value of a \oxygen[16]/\neon[20]/\magnesium[24]
mixture.  Table~\ref{tab:siessx} summarizes the abundances in the ONe
core of a representative stellar model from \citet{Siess2006}.  This
mixture has $\Ye \approx 0.49$. See their Section 5 for an explanation
of this core nucleosynthesis.

\begin{deluxetable}{lR}
  \tablenum{1}
  \tablecolumns{2}
  \tablecaption{Approximate core composition for the $Z = 0.02$ $\unit[9.5]{\Msun}$ model of \citet{Siess2006}.
    \label{tab:siessx}}
  \tablehead{
    \colhead{Isotope} & \colhead{Mass Fraction (\%)}}
  \startdata
  \carbon[12]    & \approx 1.0 \\ 
  \oxygen[16]    & \approx 55 \\ 
  \neon[20]      & \approx 30 \\
  \neon[21]\tablenotemark{a}  & \approx 0.8 \\
  \neon[22]      & \approx 0.7 \\ 
  \sodium[23]\tablenotemark{a}    & \approx 5.5 \\
  \magnesium[24]\tablenotemark{a} & \approx 3.3 \\
  \magnesium[25]\tablenotemark{a} & \approx 1.5 \\
  \magnesium[26] & \approx 0.9 \\
  \aluminum[27]\tablenotemark{a} & \approx 0.7
  \enddata
  \tablenotetext{a}{This model has a central density $\approx 7 \times \unit[10^7]{\gcc}$.  Indicated isotopes can undergo additional electron captures as the density increases towards $\approx \unit[10^{10}]{\gcc}$, the density at which the oxygen deflagration is expected to form.}
\end{deluxetable}

As the core slowly grows and its density increases
further, the Fermi energy of the
degenerate electrons rises.  Electron-capture reactions on a given
isotope become energetically favored when material exceeds its
threshold density.%
\footnote{The energetics of these weak reactions are critical for
  understanding the thermal evolution of SAGB cores and accreting ONe
  WDs \citep[e.g.,][]{Jones2013, Schwab2015,Schwab2017a}.}
In Table 1, we indicate isotopes where these electron captures are
likely to occur before the formation of the oxygen deflagration
(meaning that their threshold densities are below the threshold
density of \neon[20], which is $\approx\unit[10^{10}]{\gcc}$).  In
what follows, we focus on the most abundant of these isotopes,
\sodium[23] and \magnesium[24].  The effective threshold density for
\sodium[23] is $\approx \unit[1.6\times10^9]{\gcc}$ and for
\magnesium[24] is $\approx \unit[4\times10^9]{\gcc}$.  The timescales
for the electron capture reactions are typically shorter that the
evolutionary timescale of the object, so they are expected to
completely convert the parent isotope to its daughter.

The electron captures imply that \Ye\ spatially varies through the
core, with \Ye\ becoming lower at higher density.  By time the
deflagration forms and begins to propagate, electron captures have
already completely converted the \sodium[23] to \neon[23] and the
\magnesium[24] to \neon[24] over the inner $\approx \unit[0.2]{\Msun}$
of the star.\footnote{A representative \Ye\ profile as a function of
  mass is shown in Figure~11 of \citet{Schwab2017a}.  One caveat is
  that if a large core convection zone were to develop, as happens in
  models adopting the Schwarzschild criterion for convection
  \citep[e.g.,][]{Miyaji1980}, the central region would likely be
  homogenized.}  For the mixture in Table~\ref{tab:siessx}, this
further reduces $\Ye$ to $\approx 0.485$.

In a concluding comment, \tw\ note that $\Ye < 0.5$ is expected and
mention two calculations including reduced $\Ye$ in the form of
\neon[22].  They report that for a flame in CO with $\Ye \approx 0.498$ the speed
decreased by $\approx 5\%$ and for a flame in ONe with $\Ye = 0.48$ the
speed decreased by $\approx 30\%$.  For CO flames, the effect of
\neon[22] was studied by C07. They found the
opposite sign of the effect, with a \neon[22] mass fraction of 0.06
leading to a $\approx 30\%$ increase in the flame speed.

To quantify the effect of lower \Ye, we calculated flame speeds at
$\rho_9 = 10$ with a variable amount of neutron-rich material.  We
performed a set of calculations using each of \neon[22], \neon[23],
and \neon[24].  In all cases, the mixture had a mass fraction 0.6 of
\oxygen[16] with the remaining material being \neon[20].
Figure~\ref{fig:lowye-ne} shows the significant impact of the neutron
richness, with the flame speed relative to that at $\Ye = 0.5$
doubling by $\Ye \approx 0.488$.  The sequences with the different
isotopes overlap, indicating the speedup is largely independent of the
neutron source.

The small change in $\Ye$ does not significantly affect the internal
energy or thermal conductivity, but does lead to a significant change
in \epsnuc\ as the initial source of extra neutrons opens additional
energy producing reaction channels.  We examined the peak values of
$\epsnuc$ in the calculations shown in Figure~\ref{fig:lowye-ne} and
confirmed that the increasing flame speed is due to an increasing
\epsnuc\ at lower \Ye\ and that it quantitatively follows the
expectation from Equation~\eqref{eq:vflame}.

To illustrate that this implies a density-dependent enhancement of the
flame speed over the \tw\ result, we construct two sets of models that
initially have a mass fraction $X_{23} = 0.05$ of $A = 23$ elements
and $X_{24} = 0.05$ of $A = 24$ elements.  In one, the $A=23$ material
is always \sodium[23] and the $A=24$ material is always
\magnesium[24].  In the other, the spatially-uniform composition is
selected differently depending on the chosen $\rho_9$.  The $A=23$
material is \sodium[23] if the density is below its threshold density
and \neon[23] if it is above it, while the $A=24$ material is
\magnesium[24] if the density is below its threshold density and
\neon[24] if it is above it.  We then run these models
and extract their flame speeds.

Figure~\ref{fig:lowye} compares these two sets of calculations by
showing the ratio of the flame speed in the case where the initial
material has electron captured to the case where it has not.  Above
both threshold densities, where $\Ye$ has fallen to $\approx 0.49$,
the flame is $\approx 80\%$ faster.

\section{Fitting Formula}
\label{sec:tables}

\begin{figure}
  \includegraphics[width=\columnwidth]{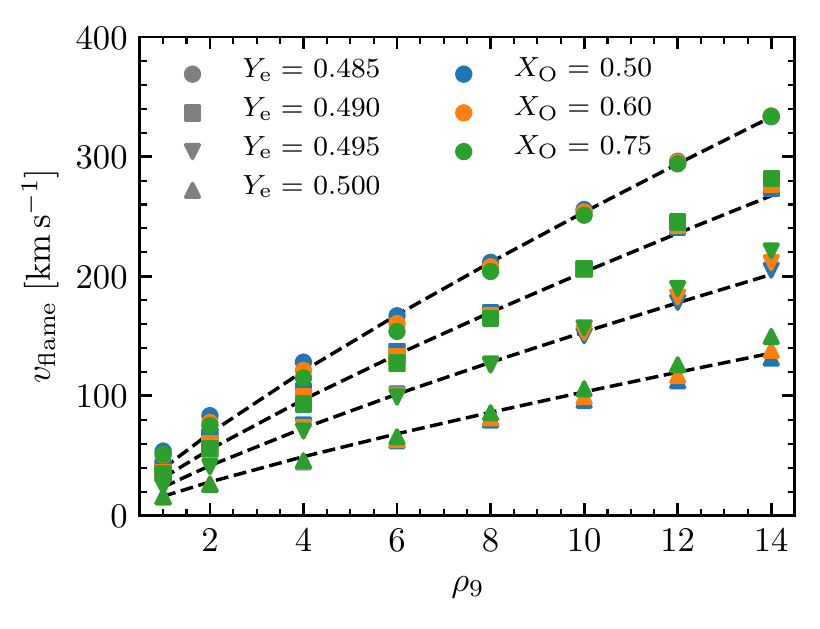}
  \caption{Flame speeds used to generate the fit given by
    Equation~\eqref{eq:fit}.  Point shape indicates $\Ye$ and point
    color indicates \Xo.  The dashed black curves show the fitting
    function.}
  \label{fig:final-v}
\end{figure}

To allow this important effect to be incorporated in hydrodynamics
calculations, we provide a simple fitting function like that of \tw,
but including $\Ye$ as an additional parameter.  As shown in
Figure~\ref{fig:lowye-ne}, the flame speed varies with $\Ye$
approximately independently of the neutron source.  Therefore, we run
a set of calculations for \oxygen[16] / \neon[20] / \neon[23]
mixtures.  We use densities $\rho_9 = \{1,2,4,6,8,10,12,14\}$,
\oxygen[16] mass fractions $\Xo = \{0.5, 0.6, 0.75\}$, and select
\neon[23] mass fractions such that
$\Ye = \{0.485, 0.490, 0.495, 0.500\}$.  Figure~\ref{fig:final-v}
plots the results.

\begin{figure}
  \includegraphics[width=\columnwidth]{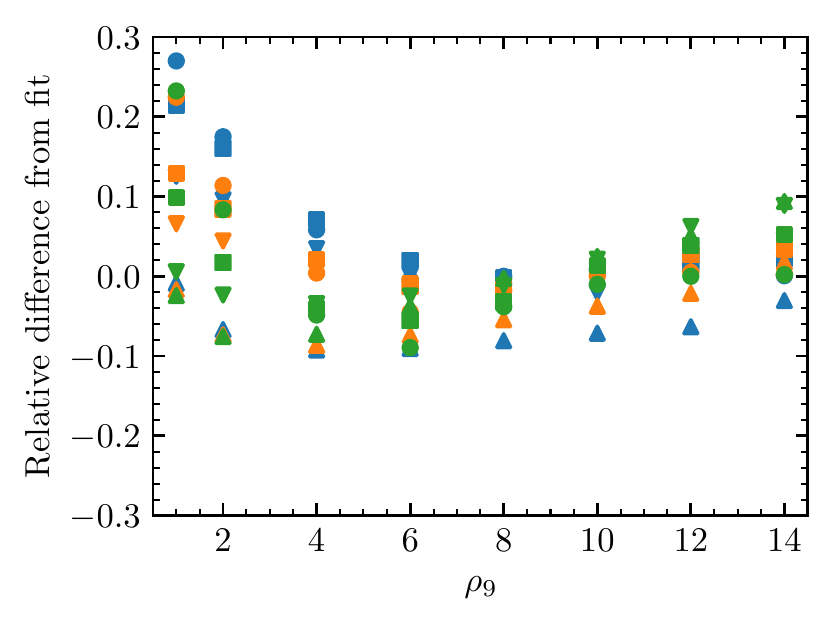}
  \caption{Relative error between calculated points and the fit.
    Point symbols are the same as in Figure~\ref{fig:final-v}. }
  \label{fig:final-err}
\end{figure}

At $\Ye = 0.5$, the flame speed always increases with increasing
oxygen abundance (as found in \tw).  However, in our results at lower
\Ye, this is no longer true.  Incorporating the effect of $\Xo$ in the
fit would require something beyond the power-law scaling used in the
fit of \tw. Given the relatively weak dependence on \Xo, we circumvent
this complication and propose the following simple fitting function
that includes only $\rho_9$ and \Ye:
\begin{equation}
  \label{eq:fit}
  v_{\rm flame}  = 16.0 \,\rho_9^{0.813} \left[1 + 96.8 \,(0.5 - \Ye)\right]\;{\rm km\,s^{-1}} ~.
\end{equation}
As shown in Figure~\ref{fig:final-err}, the fit agrees with the
calculated points within 10\% relative error at $\rho_9 > 4$, with the
maximum error growing to a 30\% underestimate at $\rho_9 = 1$.  This
fitting function will also do a worse job in pure oxygen mixtures (a
relative error $\approx 30\%$ for the $\Xo = 1$ points shown in
Figure~\ref{fig:tw92-one}), but such pure mixtures are unlikely to
arise in astrophysical contexts.  If a more precise reproduction of
our results is desired, the flame speed values are provided in Table~\ref{tab:final-grid},
allowing direct interpolation in our results.

\begin{deluxetable*}{llllllllll}
  \tablenum{2}
  \tablecolumns{10}
  \tablecaption{Flame speed (in \kms) for models described in Section~\ref{sec:tables} and shown in Figure~\ref{fig:final-v}. 
    \label{tab:final-grid}}
  \tablehead{
         \colhead{\Xo} & \colhead{\Ye} & \multicolumn{8}{c}{$\rho_9$} \\
& & \colhead{1}  & \colhead{2} & \colhead{4} & \colhead{6} & \colhead{8} & \colhead{10} & \colhead{12} & \colhead{14}}
  \startdata
0.50 & 0.485 & 53.8 & 83.5 & 128 & 167 & 212 & 256 & 296 & 333 \\
0.50 & 0.490 & 40.1 & 65.8 & 104 & 137 & 170 & 206 & 241 & 273 \\
0.50 & 0.495 & 27.2 & 46.1 & 75.6 & 102 & 126 & 150 & 178 & 205 \\
0.50 & 0.500 & 15.9 & 26.3 & 45.0 & 62.6 & 79.8 & 96.4 & 113 & 132 \\
0.60 & 0.485 & 50.7 & 77.7 & 121 & 160 & 208 & 254 & 295 & 334 \\
0.60 & 0.490 & 36.2 & 60.3 & 98.9 & 133 & 167 & 206 & 242 & 277 \\
0.60 & 0.495 & 25.4 & 43.6 & 73.5 & 100 & 126 & 153 & 182 & 211 \\
0.60 & 0.500 & 15.7 & 26.1 & 45.3 & 63.6 & 81.8 & 99.6 & 117 & 138 \\
0.75 & 0.485 & 51.2 & 75.1 & 115 & 154 & 204 & 251 & 294 & 334 \\
0.75 & 0.490 & 35.0 & 56.2 & 93.1 & 127 & 165 & 206 & 245 & 282 \\
0.75 & 0.495 & 23.9 & 40.7 & 70.6 & 99.0 & 127 & 157 & 190 & 221 \\
0.75 & 0.500 & 15.6 & 26.1 & 45.9 & 65.8 & 86.0 & 106 & 126 & 150 \\
\enddata
\tablecomments{A machine readable version of this data is provided.}
\end{deluxetable*}

\section{Summary and Conclusions}
\label{sec:conclusions}

Using \MESA\ calculations that resolve the structure of
conductively-propagating deflagrations, we calculated laminar flame
speeds in oxygen-neon mixtures.  These speeds are a necessary
ingredient in simulations of the final stages of electron-capture
supernovae and accretion-induced collapse.

We demonstrated that the values of $\Ye \approx 0.48-0.49$ expected in
these objects lead to an increase in the flame speed by a factor of
$\approx 2$ over that at $\Ye = 0.5$, the value assumed in
the widely-used prescription of \citet{Timmes1992}.
The low $\Ye$ is due to the nucleosynthesis during the helium and carbon
burning phases that preceded the formation of the ONe core and to
subsequent electron captures on isotopes initially present in the ONe
core (most importantly \sodium[23] and \magnesium[24]) that occur as
the core grows.  As shown in Figure~\ref{fig:lowye}, this implies that
the realized enhancement is density-dependent and most significant
for $\rho \gtrsim \unit[4\times10^9]{\gcc}$ (i.e., above the
\magnesium[24] threshold density).

Full star hydrodynamics simulations that follow the propagation of the
deflagration through the ONe core do so by including a sub-grid model
for the flame.  These models enhance the laminar speed by including a
sub-grid model of the flame-turbulence interaction (which allows for a
larger, non-planar area to undergo burning), such that the laminar
speed is only a lower limit. Eventually, this speed becomes irrelevant
once the turbulence is fully developed, as a turbulent deflagration no longer
depends on the laminar speed.  In Section~\ref{sec:tables}, we
provide a tabulated set of laminar flame speeds as well as a
convenient fitting function.  These are suitable for incorporation
into sub-grid flame models.

Figures 4 and 5 in \citet{Jones2016c} show the laminar and turbulent
flame velocities in their 3D hydrodynamic simulations.  Typically,
these flames remain laminar for $\approx \unit[0.4]{s}$, corresponding
to $\approx \unit[100]{km}$ of flame propagation.  Typically, the inner
$\approx \unit[200]{km}$ is above the \magnesium[24] threshold density
and thus at the lowest \Ye.  Therefore, we believe the factor of 2
speedup is representative of what will be realized in practice.  The
more rapid release of energy associated with a faster flame pushes
models in the direction of being more likely to explode (meaning less
likely to form a neutron star).  The full implications of our results
await the incorporation of this updated prescription in
multidimensional models.

\acknowledgments

We thank Sam Jones, Pablo Marchant, Bill Paxton, and Fritz R\"{o}pke for helpful conversations.
We thank the referee for a useful report.
This work benefited from the May 2019 Lorentz Center program
Electron-Capture-Initiated Stellar Collapse.  We are grateful to Zo\"{e}
Weber-Porter for performing some exploratory calculations of this
problem as part of her UC Santa Cruz undergraduate thesis.
We acknowledge use of the lux supercomputer at UC Santa Cruz, funded by NSF MRI grant AST 1828315.
We thank Brant Robertson for the rapid and friendly technical support that enabled us to make use of this machine.
This research  was supported by the National Science Foundation (NSF) under the Software Infrastructure 
for Sustained Innovation program grants (ACI-1663684, ACI-1663688, ACI-1663696).
JS is supported by the A.F. Morrison Fellowship in Lick Observatory.
RF is supported by the Netherlands Organization for Scientific Research (NWO) 
through a top module 2 grant with project number 614.001.501 (PI de Mink). 
This research was also supported at ASU by the NSF under grant PHY-1430152 for the 
Physics Frontier Center “Joint Institute for Nuclear Astrophysics—Center for the Evolution of the Elements” (JINA-CEE).
This research made extensive use of the SAO/NASA Astrophysics Data System (ADS).

\software{
\MESA \citep{Paxton2011, Paxton2013, Paxton2015, Paxton2018, Paxton2019},
\texttt{MESASDK} 20190830 \citep{mesasdk}, 
\texttt{sig99} \ \url{http://cococubed.asu.edu/code_pages/kap.shtml},
\texttt{matplotlib} \citep{hunter2007}, 
\texttt{NumPy} \citep{walt2011},
\texttt{py\_mesa\_reader} \citep{pmr},
\texttt{MesaScript} \citep{MesaScript}
}

\appendix

\section{Convergence Studies}
\label{sec:convergence}

\begin{figure}
  \gridline{
    \fig{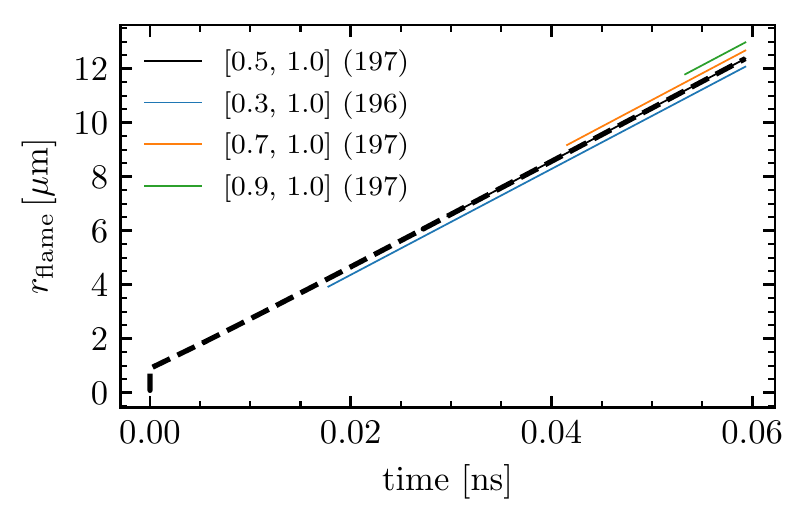}{0.5\textwidth}{(a) Varying the portion of data used to measure flame speed}
    \fig{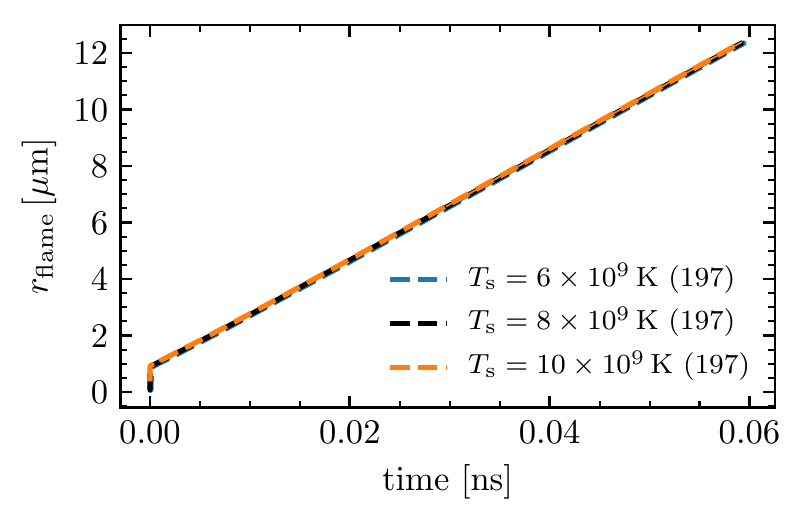}{0.5\textwidth}{(b) Varying the temperature of the initial hot spot \Ts}
  }
  \gridline{
    \fig{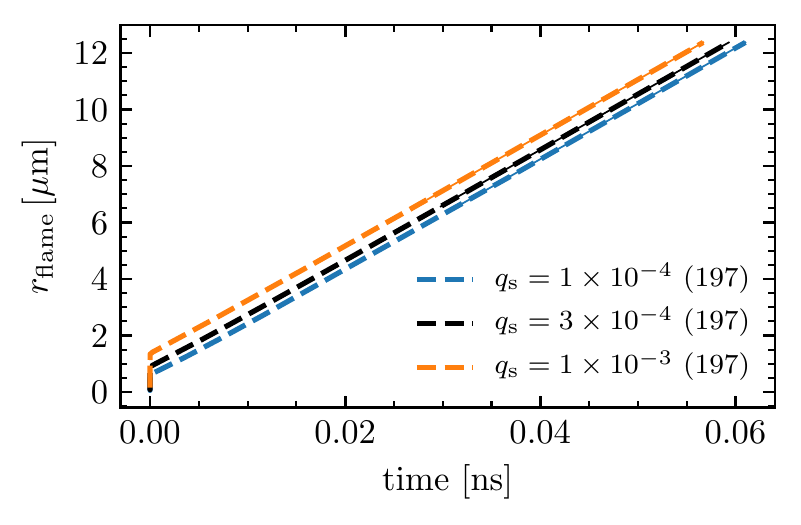}{0.5\textwidth}{(c) Varying the size of the initial hotspot \qs}
    \fig{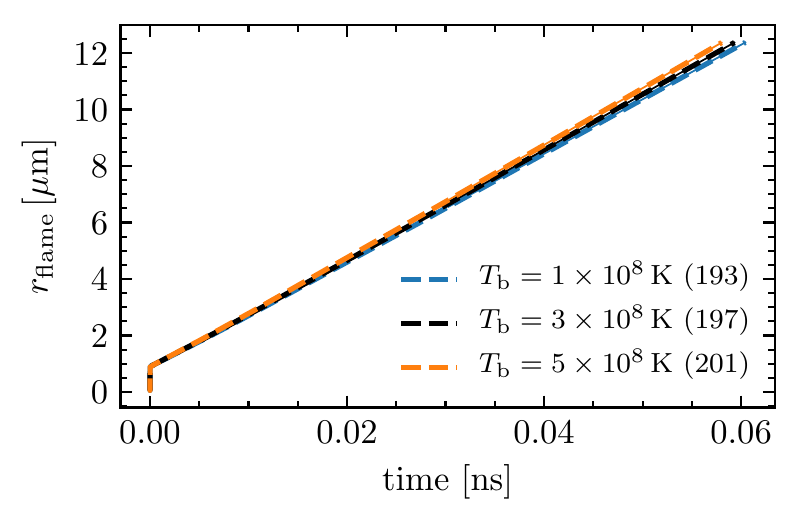}{0.5\textwidth}{(d) Varying the upstream temperature \Tb}
  }
  \gridline{
    \fig{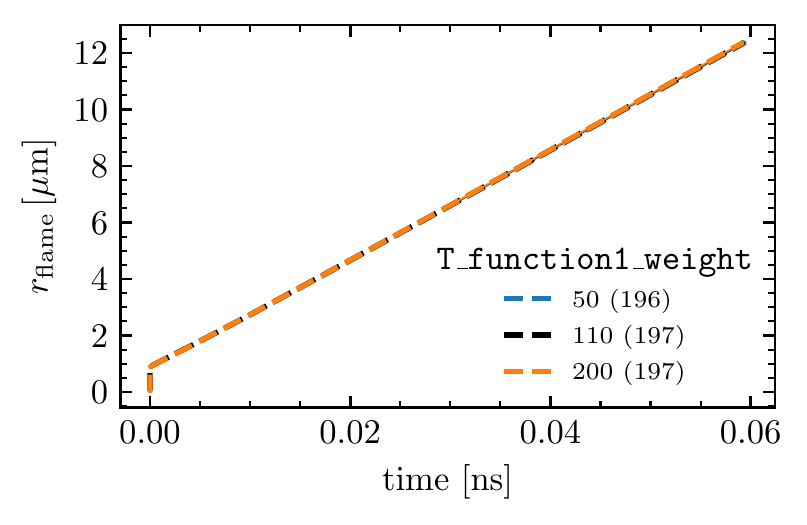}{0.5\textwidth}{(e) Varying spatial resolution}
    \fig{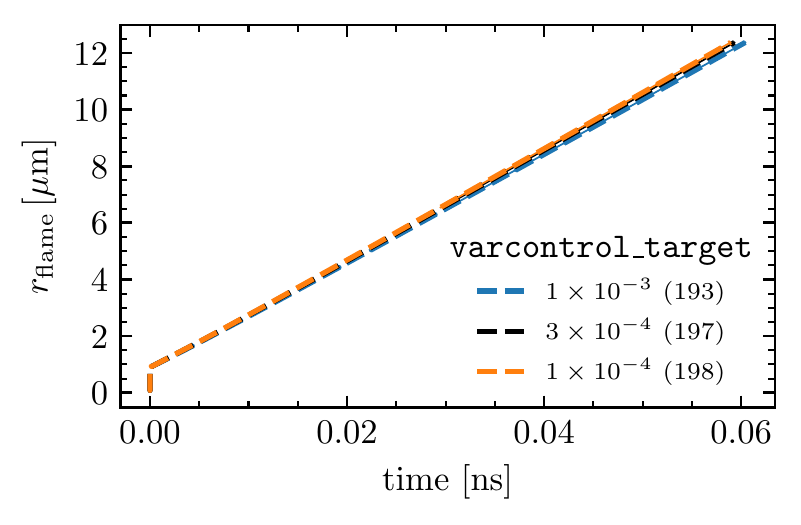}{0.5\textwidth}{(f) Varying temporal resolution}
  }
  \caption{Effect of modeling choices on flame speeds for the
    fiducial CO case $(\Xc = 0.5, \rho_9 = 6$).  Default choices are indicated with black lines.
    The value in parentheses in the legend is the flame speed in \kms.}
  \label{fig:convergence-co}
\end{figure}

\begin{figure}
  \gridline{
    \fig{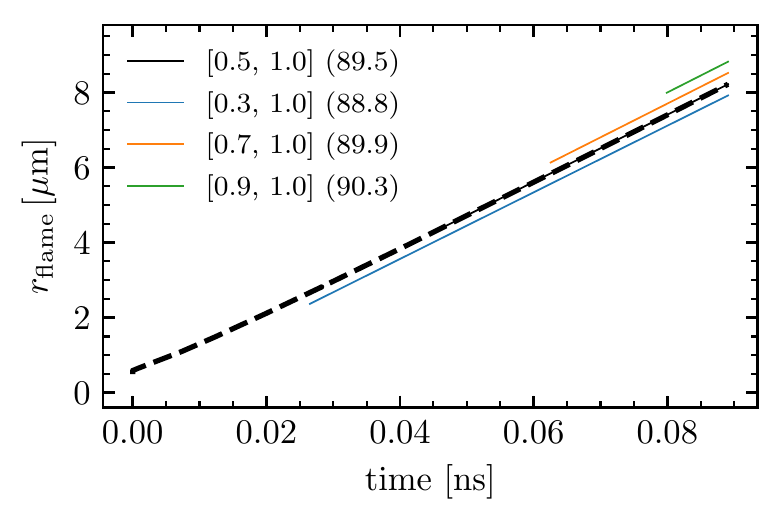}{0.5\textwidth}{(a) Varying the portion of data used to measure flame speed}
    \fig{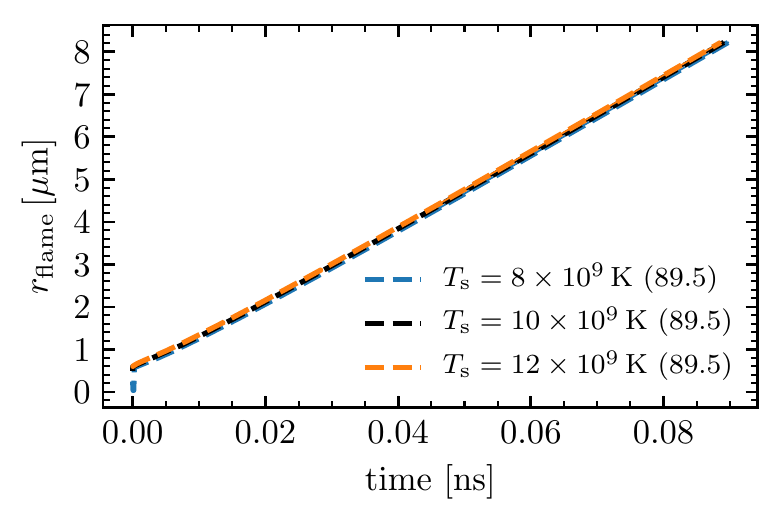}{0.5\textwidth}{(b) Varying the temperature of the initial hot spot \Ts}
  }
  \gridline{
    \fig{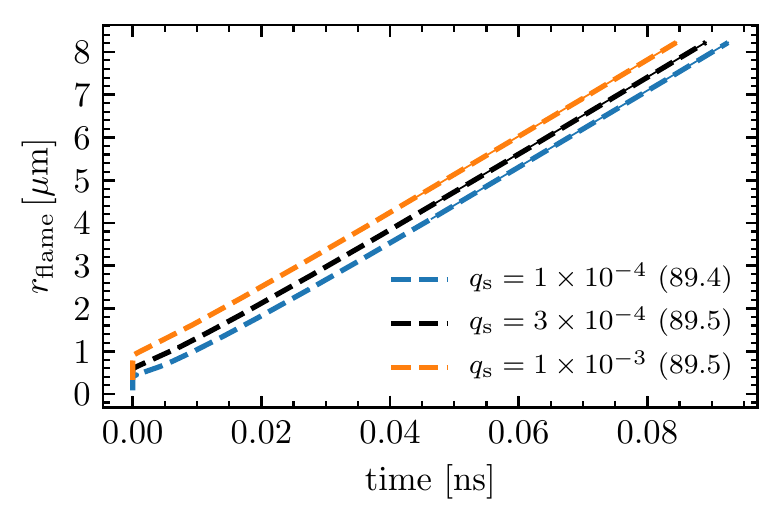}{0.5\textwidth}{(c) Varying the size of the initial hotspot \qs}
    \fig{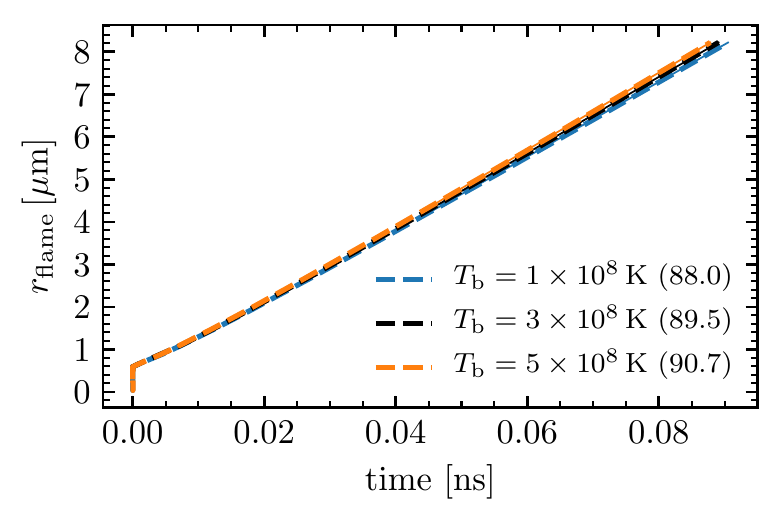}{0.5\textwidth}{(d) Varying the upstream temperature \Tb}
  }
  \gridline{
    \fig{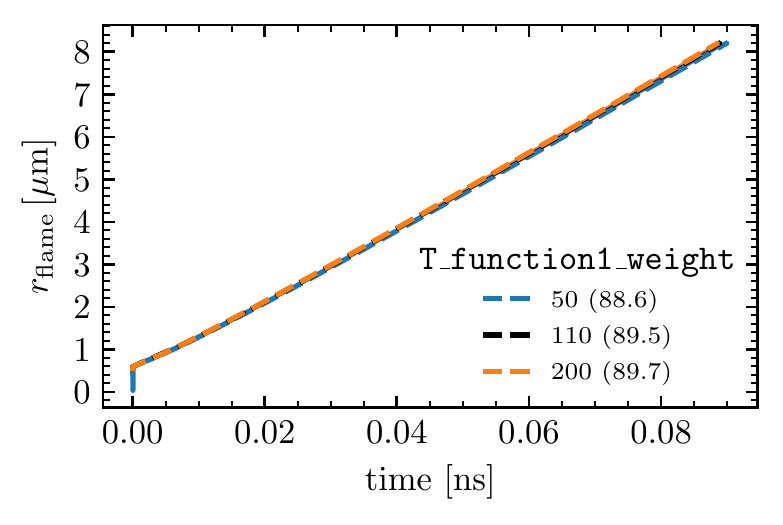}{0.5\textwidth}{(e) Varying spatial resolution}
    \fig{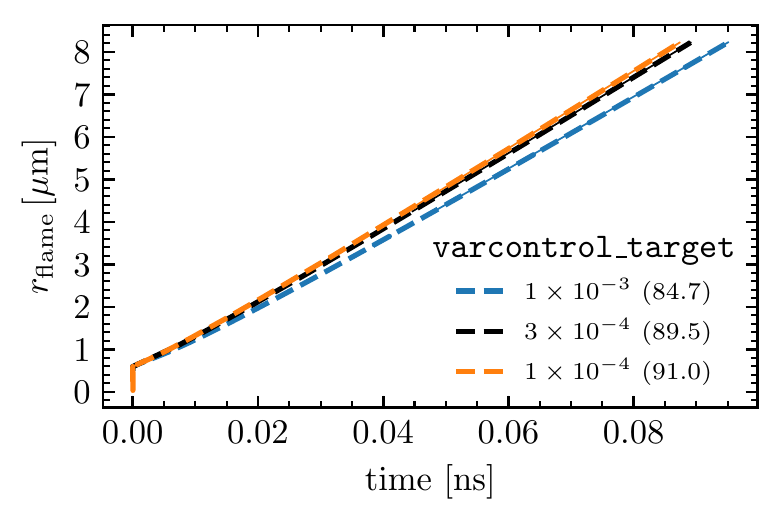}{0.5\textwidth}{(f) Varying temporal resolution}
  }
  \caption{Effect of modeling choices on flame speed for the
    fiducial ONe case $(\Xo = 0.6, \rho_9 = 10$).  Default choices are indicated with black lines.
    The value in parentheses in the legend is the flame speed in \kms.}
  \label{fig:convergence-one}
\end{figure}

In this appendix, we demonstrate that the flame speeds we report are
only weakly dependent on the details of the initial conditions and the
spatial and temporal resolution of the \MESA\ calculations.
Figure~\ref{fig:convergence-co} shows this for a flame in a CO mixture
and Figure~\ref{fig:convergence-one} does so for a flame in an ONe
mixture.  The discussion below applies equally to both figures.

Panel (a) illustrates the effect of varying our procedure for
extracting the flame speed.  By default, we measure the flame
speed using the change in position over the second half of the time
interval covered by the calculation (indicated as [0.5, 1.0] in the
legend).  So long as we avoid the transients during the early part (roughly the first
quarter) of the calculation, the extracted velocities are consistent
at the percent level.

Panel (b) and panel (c) illustrate the effect of varying the
temperature of the initial hot spot \Ts\ and its fractional size \qs.
As expected, so long as the hot spot causes a steady-state flame to propagate, these
choices have no effect on the flame speed.

Panel (d) illustrates the effect of varying the upstream temperature
\Tb.  The flame speed increases with increasing upstream
temperature, but such that a factor of $\approx 2$ change in \Tb\
leads to only $1-2\%$ increase in the flame speed.  We would
expect this to remain true so long as $\Tb$ is much less than the
post-burn temperature of $\approx \unit[10^{10}]{K}$.

Panel (e) illustrates the effect of varying the spatial resolution of
the \MESA\ calculation.  \MESA\ adaptively refines its mesh based on a
set of mesh functions.  The maximum cell-to-cell variation of these
functions is maintained at around the value of the control
\texttt{mesh\_delta\_coeff} which is set equal to 1 in our
calculations.  One of the built-in mesh functions has the form
$\texttt{T\_function1\_weight}\times\log(T/\rm K)$.  This function
ensures that temperature gradients are resolved, placing approximately
\texttt{T\_function1\_weight} zones per dex change in temperature.
The number of zones in the calculation (which is $\approx 1000$ for
the default) varies roughly linearly with
\texttt{T\_function1\_weight}.  The results are approximately
independent of the spatial resolution, with a sub-percent increase
between the default and higher resolution cases.

Panel (f) illustrates the effect of varying the temporal resolution of
the \MESA\ calculation.  The control \texttt{varcontrol\_target}
limits the fractional step-to-step variation of quantities in the same
cell.  The number of timesteps in the calculation (which is
$\approx 2000$ for the default) varies roughly linearly with the
inverse of \texttt{varcontrol\_target}.  The results are approximately
independent of the time resolution, with a roughly 1\% increase
between the default and the highest resolution.

\clearpage
\bibliographystyle{aasjournal}
\bibliography{flames.bib}

\end{document}

%% file: macros.tex
\newcommand{\tw}{TW92}

\newcommand{\MESA}{{\tt MESA}}

\newcommand{\Msun}{\ensuremath{\mathrm{M}_\odot}}
\newcommand{\gcc}{\ensuremath{\mathrm{g\,cm^{-3}}}} 

\newcommand{\Ye}{\ensuremath{Y_{\rm e}}}

\newcommand{\Tb}{\ensuremath{T_{\rm b}}}

\newcommand{\Ts}{\ensuremath{T_{\rm s}}}
\newcommand{\qs}{\ensuremath{q_{\rm s}}}

\newcommand{\kms}{\ensuremath{\rm km\,s^{-1}}}

\newcommand{\epsnuc}{\ensuremath{\epsilon_{\mathrm{nuc}}}}
\newcommand{\kappacond}{\ensuremath{\kappa_{\mathrm{cond}}}}

\newcommand{\Xc}{\ensuremath{X_{\mathrm{C}}}}
\newcommand{\Xo}{\ensuremath{X_{\mathrm{O}}}}
\newcommand{\Xne}{\ensuremath{X_{\mathrm{Ne}}}}

%% file: nuclides.tex
\newcommand{\nuclei}[2]{\ensuremath{\mathrm{^{#1}#2}}}

\newcommand{\carbon}[1][12]{\nuclei{#1}{C}}

\newcommand{\oxygen}[1][16]{\nuclei{#1}{O}}

\newcommand{\neon}[1][20]{\nuclei{#1}{Ne}}
\newcommand{\sodium}[1][23]{\nuclei{#1}{Na}}
\newcommand{\magnesium}[1][24]{\nuclei{#1}{Mg}}
\newcommand{\aluminum}[1][27]{\nuclei{#1}{Al}}


%% file: flames.bbl
\begin{thebibliography}{}
\expandafter\ifx\csname natexlab\endcsname\relax\def\natexlab#1{#1}\fi
\providecommand{\url}[1]{\href{#1}{#1}}
\providecommand{\dodoi}[1]{doi:~\href{http://doi.org/#1}{\nolinkurl{#1}}}
\providecommand{\doeprint}[1]{\href{http://ascl.net/#1}{\nolinkurl{http://ascl.net/#1}}}
\providecommand{\doarXiv}[1]{\href{https://arxiv.org/abs/#1}{\nolinkurl{https://arxiv.org/abs/#1}}}

\bibitem[{{Alastuey} \& {Jancovici}(1978)}]{Alastuey1978}
{Alastuey}, A., \& {Jancovici}, B. 1978, \apj, 226, 1034,
  \dodoi{10.1086/156681}

\bibitem[{{Angulo} {et~al.}(1999){Angulo}, {Arnould}, {Rayet}, {Descouvemont},
  {Baye}, {Leclercq-Willain}, {Coc}, {Barhoumi}, {Aguer}, {Rolfs}, {Kunz},
  {Hammer}, {Mayer}, {Paradellis}, {Kossionides}, {Chronidou}, {Spyrou},
  {degl'Innocenti}, {Fiorentini}, {Ricci}, {Zavatarelli}, {Providencia},
  {Wolters}, {Soares}, {Grama}, {Rahighi}, {Shotter}, \& {Lamehi
  Rachti}}]{angulo99}
{Angulo}, C., {Arnould}, M., {Rayet}, M., {et~al.} 1999, \nphysa, 656, 3,
  \dodoi{10.1016/S0375-9474(99)00030-5}

\bibitem[{{Brooks} {et~al.}(2017){Brooks}, {Schwab}, {Bildsten}, {Quataert},
  {Paxton}, {Blinnikov}, \& {Sorokina}}]{Brooks2017d}
{Brooks}, J., {Schwab}, J., {Bildsten}, L., {et~al.} 2017, \apj, 850, 127,
  \dodoi{10.3847/1538-4357/aa9568}

\bibitem[{{Canal} {et~al.}(1992){Canal}, {Isern}, \& {Labay}}]{Canal1992}
{Canal}, R., {Isern}, J., \& {Labay}, J. 1992, \apjl, 398, L49,
  \dodoi{10.1086/186574}

\bibitem[{{Canal} \& {Schatzman}(1976)}]{Canal1976}
{Canal}, R., \& {Schatzman}, E. 1976, \aap, 46, 229

\bibitem[{{Chamulak} {et~al.}(2007){Chamulak}, {Brown}, \&
  {Timmes}}]{Chamulak2007}
{Chamulak}, D.~A., {Brown}, E.~F., \& {Timmes}, F.~X. 2007, \apjl, 655, L93,
  \dodoi{10.1086/511856}

\bibitem[{{Chugunov} {et~al.}(2007){Chugunov}, {Dewitt}, \&
  {Yakovlev}}]{Chugunov2007}
{Chugunov}, A.~I., {Dewitt}, H.~E., \& {Yakovlev}, D.~G. 2007, \prd, 76,
  025028, \dodoi{10.1103/PhysRevD.76.025028}

\bibitem[{{Cyburt} {et~al.}(2010){Cyburt}, {Amthor}, {Ferguson}, {Meisel},
  {Smith}, {Warren}, {Heger}, {Hoffman}, {Rauscher}, {Sakharuk}, {Schatz},
  {Thielemann}, \& {Wiescher}}]{Cyburt2010}
{Cyburt}, R.~H., {Amthor}, A.~M., {Ferguson}, R., {et~al.} 2010, \apjs, 189,
  240, \dodoi{10.1088/0067-0049/189/1/240}

\bibitem[{{Dewitt} {et~al.}(1973){Dewitt}, {Graboske}, \&
  {Cooper}}]{Dewitt1973}
{Dewitt}, H.~E., {Graboske}, H.~C., \& {Cooper}, M.~S. 1973, \apj, 181, 439,
  \dodoi{10.1086/152061}

\bibitem[{{Farmer} {et~al.}(2015){Farmer}, {Fields}, \& {Timmes}}]{Farmer2015}
{Farmer}, R., {Fields}, C.~E., \& {Timmes}, F.~X. 2015, \apj, 807, 184,
  \dodoi{10.1088/0004-637X/807/2/184}

\bibitem[{{Garcia-Berro} {et~al.}(1997){Garcia-Berro}, {Ritossa}, \&
  {Iben}}]{GarciaBerro1997b}
{Garcia-Berro}, E., {Ritossa}, C., \& {Iben}, Jr., I. 1997, \apj, 485, 765,
  \dodoi{10.1086/304444}

\bibitem[{{Graboske} {et~al.}(1973){Graboske}, {Dewitt}, {Grossman}, \&
  {Cooper}}]{Graboske1973}
{Graboske}, H.~C., {Dewitt}, H.~E., {Grossman}, A.~S., \& {Cooper}, M.~S. 1973,
  \apj, 181, 457, \dodoi{10.1086/152062}

\bibitem[{Hunter(2007)}]{hunter2007}
Hunter, J.~D. 2007, Computing In Science \&amp; Engineering, 9, 90

\bibitem[{{Iben} {et~al.}(1997){Iben}, {Ritossa}, \& {Garcia-Berro}}]{Iben1997}
{Iben}, Jr., I., {Ritossa}, C., \& {Garcia-Berro}, E. 1997, \apj, 489, 772,
  \dodoi{10.1086/304822}

\bibitem[{{Itoh} {et~al.}(1979){Itoh}, {Totsuji}, {Ichimaru}, \&
  {Dewitt}}]{Itoh1979}
{Itoh}, N., {Totsuji}, H., {Ichimaru}, S., \& {Dewitt}, H.~E. 1979, \apj, 234,
  1079, \dodoi{10.1086/157590}

\bibitem[{{Jones} {et~al.}(2016){Jones}, {R{\"o}pke}, {Pakmor}, {Seitenzahl},
  {Ohlmann}, \& {Edelmann}}]{Jones2016c}
{Jones}, S., {R{\"o}pke}, F.~K., {Pakmor}, R., {et~al.} 2016, \aap, 593, A72,
  \dodoi{10.1051/0004-6361/201628321}

\bibitem[{{Jones} {et~al.}(2013){Jones}, {Hirschi}, {Nomoto}, {Fischer},
  {Timmes}, {Herwig}, {Paxton}, {Toki}, {Suzuki}, {Mart{\'{\i}}nez-Pinedo},
  {Lam}, \& {Bertolli}}]{Jones2013}
{Jones}, S., {Hirschi}, R., {Nomoto}, K., {et~al.} 2013, \apj, 772, 150,
  \dodoi{10.1088/0004-637X/772/2/150}

\bibitem[{{Jones} {et~al.}(2019){Jones}, {R{\"o}pke}, {Fryer}, {Ruiter},
  {Seitenzahl}, {Nittler}, {Ohlmann}, {Reifarth}, {Pignatari}, \&
  {Belczynski}}]{Jones2019a}
{Jones}, S., {R{\"o}pke}, F.~K., {Fryer}, C., {et~al.} 2019, \aap, 622, A74,
  \dodoi{10.1051/0004-6361/201834381}

\bibitem[{{Leung} {et~al.}(2019){Leung}, {Nomoto}, \& {Suzuki}}]{Leung2019b}
{Leung}, S.-C., {Nomoto}, K., \& {Suzuki}, T. 2019, arXiv e-prints,
  arXiv:1901.11438.
\newblock \doarXiv{1901.11438}

\bibitem[{{Miyaji} \& {Nomoto}(1987)}]{Miyaji1987}
{Miyaji}, S., \& {Nomoto}, K. 1987, \apj, 318, 307, \dodoi{10.1086/165368}

\bibitem[{{Miyaji} {et~al.}(1980){Miyaji}, {Nomoto}, {Yokoi}, \&
  {Sugimoto}}]{Miyaji1980}
{Miyaji}, S., {Nomoto}, K., {Yokoi}, K., \& {Sugimoto}, D. 1980, \pasj, 32, 303

\bibitem[{{Nomoto} \& {Kondo}(1991)}]{Nomoto1991}
{Nomoto}, K., \& {Kondo}, Y. 1991, \apjl, 367, L19, \dodoi{10.1086/185922}

\bibitem[{Paxton(2019)}]{MESA}
Paxton, B. 2019, {Modules for Experiments in Stellar Astrophysics (MESA)},
  r12115,  Zenodo, \dodoi{10.5281/zenodo.3473377}.
\newblock \url{https://doi.org/10.5281/zenodo.3473377}

\bibitem[{{Paxton} {et~al.}(2011){Paxton}, {Bildsten}, {Dotter}, {Herwig},
  {Lesaffre}, \& {Timmes}}]{Paxton2011}
{Paxton}, B., {Bildsten}, L., {Dotter}, A., {et~al.} 2011, \apjs, 192, 3,
  \dodoi{10.1088/0067-0049/192/1/3}

\bibitem[{{Paxton} {et~al.}(2013){Paxton}, {Cantiello}, {Arras}, {Bildsten},
  {Brown}, {Dotter}, {Mankovich}, {Montgomery}, {Stello}, {Timmes}, \&
  {Townsend}}]{Paxton2013}
{Paxton}, B., {Cantiello}, M., {Arras}, P., {et~al.} 2013, \apjs, 208, 4,
  \dodoi{10.1088/0067-0049/208/1/4}

\bibitem[{{Paxton} {et~al.}(2015){Paxton}, {Marchant}, {Schwab}, {Bauer},
  {Bildsten}, {Cantiello}, {Dessart}, {Farmer}, {Hu}, {Langer}, {Townsend},
  {Townsley}, \& {Timmes}}]{Paxton2015}
{Paxton}, B., {Marchant}, P., {Schwab}, J., {et~al.} 2015, \apjs, 220, 15,
  \dodoi{10.1088/0067-0049/220/1/15}

\bibitem[{{Paxton} {et~al.}(2018){Paxton}, {Schwab}, {Bauer}, {Bildsten},
  {Blinnikov}, {Duffell}, {Farmer}, {Goldberg}, {Marchant}, {Sorokina},
  {Thoul}, {Townsend}, \& {Timmes}}]{Paxton2018}
{Paxton}, B., {Schwab}, J., {Bauer}, E.~B., {et~al.} 2018, \apjs, 234, 34,
  \dodoi{10.3847/1538-4365/aaa5a8}

\bibitem[{{Paxton} {et~al.}(2019){Paxton}, {Smolec}, {Schwab}, {Gautschy},
  {Bildsten}, {Cantiello}, {Dotter}, {Farmer}, {Goldberg}, {Jermyn}, {Kanbur},
  {Marchant}, {Thoul}, {Townsend}, {Wolf}, {Zhang}, \& {Timmes}}]{Paxton2019}
{Paxton}, B., {Smolec}, R., {Schwab}, J., {et~al.} 2019, \apjs, 243, 10,
  \dodoi{10.3847/1538-4365/ab2241}

\bibitem[{{Poelarends} {et~al.}(2017){Poelarends}, {Wurtz}, {Tarka}, {Cole
  Adams}, \& {Hills}}]{Poelarends2017}
{Poelarends}, A.~J.~T., {Wurtz}, S., {Tarka}, J., {Cole Adams}, L., \& {Hills},
  S.~T. 2017, \apj, 850, 197, \dodoi{10.3847/1538-4357/aa988a}

\bibitem[{{Saio} \& {Nomoto}(1985)}]{Saio1985b}
{Saio}, H., \& {Nomoto}, K. 1985, \aap, 150, L21

\bibitem[{{Schwab} {et~al.}(2017){Schwab}, {Bildsten}, \&
  {Quataert}}]{Schwab2017a}
{Schwab}, J., {Bildsten}, L., \& {Quataert}, E. 2017, \mnras, 472, 3390,
  \dodoi{10.1093/mnras/stx2169}

\bibitem[{{Schwab} {et~al.}(2015){Schwab}, {Quataert}, \&
  {Bildsten}}]{Schwab2015}
{Schwab}, J., {Quataert}, E., \& {Bildsten}, L. 2015, \mnras, 453, 1910,
  \dodoi{10.1093/mnras/stv1804}

\bibitem[{{Schwab} \& {Rocha}(2019)}]{Schwab2019}
{Schwab}, J., \& {Rocha}, K.~A. 2019, \apj, 872, 131,
  \dodoi{10.3847/1538-4357/aaffdc}

\bibitem[{{Siess}(2006)}]{Siess2006}
{Siess}, L. 2006, \aap, 448, 717, \dodoi{10.1051/0004-6361:20053043}

\bibitem[{{Tauris} {et~al.}(2015){Tauris}, {Langer}, \&
  {Podsiadlowski}}]{Tauris2015b}
{Tauris}, T.~M., {Langer}, N., \& {Podsiadlowski}, P. 2015, \mnras, 451, 2123,
  \dodoi{10.1093/mnras/stv990}

\bibitem[{{Timmes}(2000)}]{Timmes2000a}
{Timmes}, F.~X. 2000, \apj, 528, 913, \dodoi{10.1086/308203}

\bibitem[{{Timmes} \& {Woosley}(1992)}]{Timmes1992}
{Timmes}, F.~X., \& {Woosley}, S.~E. 1992, \apj, 396, 649,
  \dodoi{10.1086/171746}

\bibitem[{Townsend(2019)}]{mesasdk}
Townsend, R. 2019, MESA SDK for Linux, 20190830,  Zenodo,
  \dodoi{10.5281/zenodo.3560834}.
\newblock \url{https://doi.org/10.5281/zenodo.3560834}

\bibitem[{van~der Walt {et~al.}(2011)van~der Walt, Colbert, \&
  Varoquaux}]{walt2011}
van~der Walt, S., Colbert, S.~C., \& Varoquaux, G. 2011, Computing in Science
  Engineering, 13, 22, \dodoi{10.1109/MCSE.2011.37}

\bibitem[{Wolf {et~al.}(2017)Wolf, Bauer, \& Schwab}]{MesaScript}
Wolf, B., Bauer, E.~B., \& Schwab, J. 2017, {wmwolf/MesaScript: A DSL for
  Writing MESA Inlists}, \dodoi{10.5281/zenodo.826954}.
\newblock \url{https://doi.org/10.5281/zenodo.826954}

\bibitem[{Wolf \& Schwab(2017)}]{pmr}
Wolf, B., \& Schwab, J. 2017, wmwolf/py\_mesa\_reader: Interact with MESA
  Output, \dodoi{10.5281/zenodo.826958}.
\newblock \url{https://doi.org/10.5281/zenodo.826958}

\end{thebibliography}
